\documentclass[aps,amsmath,amssymb,reprint, double column,prl, showkeys]{revtex4-2}

\usepackage{graphicx,epsfig}
\usepackage{amssymb}
\usepackage{amsmath}
\usepackage{bm}
\usepackage{textcomp}
\usepackage{color}

\usepackage{soul}

\begin{document}
\setcounter{page}{1}

\title{Competition between fractional quantum Hall liquid and Wigner solid at small fillings: Role of layer thickness and Landau level mixing}
\author{K. A. \surname{Villegas Rosales}}
\author{S. K. \surname{Singh}}
\author{Meng K. \surname{Ma}}
\author{Md. Shafayat \surname{Hossain}}
\author{Y. J. \surname{Chung}}
\author{L. N. \surname{Pfeiffer}}
\author{K. W. \surname{West}}
\author{K. W. \surname{Baldwin}}
\author{M. \surname{Shayegan}}
\affiliation{Department of Electrical Engineering, Princeton University, Princeton, New Jersey 08544, USA}
\date{\today}

\begin{abstract}

What is the fate of the ground state of a two-dimensional electron system (2DES) at very low Landau level filling factors ($\nu$) where interaction reigns supreme? An ordered array of electrons, the so-called Wigner crystal, has long been believed to be the answer. It was in fact the search for the elusive Wigner crystal that led to the discovery of an unexpected, incompressible liquid state, namely the fractional quantum Hall state at $\nu=1/3$. Understanding the competition between the liquid and solid ground states has since remained an active field of fundamental research. Here we report experimental data for a new two-dimensional system where the electrons are confined to an AlAs quantum well. The exceptionally high quality of the samples and the large electron effective mass allow us to determine the liquid-solid phase diagram for the two-dimensional electrons in a large range of filling factors near $\simeq 1/3$ and $\simeq 1/5$. The data and their comparison with an available theoretical phase diagram reveal the crucial role of Landau level mixing and finite electron layer thickness in determining the prevailing ground states.

\end{abstract}

\maketitle

The ground states of an interacting, low-disorder two-dimensional electron system (2DES) in a large perpendicular magnetic field ($B$) has been of continued interest for decades. This is a regime where the kinetic (Fermi) energy of the electrons is quenched as they all occupy the highly-degenerate, lowest Landau level (LL), and the interaction (Coulomb) energy dominates. As the 2DES enters deep into this extreme quantum limit, the LL filling factor ($\nu=nh/eB$) becomes very small; $n$ is the 2DES density. At sufficiently small fillings $\nu\ll1$, the ground state should be a magnetic-field-induced Wigner solid, where the electrons arrange themselves in an ordered (triangular) array to minimize their mutual Coulomb potential energy \cite{Wigner.PR.1934, Lozovik.JTEP.1975}. Experiments on a relatively low-disorder GaAs 2DES, however, revealed a new phase of matter at $\nu=1/3$, namely the fractional quantum Hall state (FQHS), an incompressible liquid state with a vanishing longitudinal resistance and a quantized Hall resistance \cite{Tsui.PRL.1982}. The search for the WS then moved to even lower fillings, and culminated in the observation of insulating phases flanking a well-developed FQHS at $\nu=1/5$ in very high-quality 2DESs \cite{Jiang.PRL.1990, Goldman.PRL.1990}. The non-linear $I$-$V$ characteristics of these insulating phases \cite{Goldman.PRL.1990}, as well as resonances in their microwave spectrum \cite{Andrei.PRL.1988} were interpreted as evidence for the formation of a WS, which is pinned by the ubiquitous disorder that is present in all real samples. This interpretation has been corroborated by numerous theoretical calculations \cite{Lam.PRB.1984, Levesque.PRB.1984, Fertig.Review.Das.Sarma&Piczuk.Book.1996, Archer&Jain.PRL.2013}, and a variety of experiments \cite{Li&Sajoto.PRL.1991, Goldys&Foxon.PRB.1992, Kukushkin.Europhys.1993, Ye.PRL.2002, Pan&Tsui.PRL.2002, Chen&Engel.NatPhys.2006, Tiemann&Muraki.NatPhys.2014, Deng&Shayegan.PRL.2016, Jang&Ashoori.NatPhys.2017, Deng&Shayegan.PRL.2019,Maryenko.NatComm.2018}.

Shortly after the observation of insulating phases near $\nu=1/5$ in GaAs 2DESs \cite{Jiang.PRL.1990, Goldman.PRL.1990}, Santos \textit{et al.} reported similar insulating phases in a dilute GaAs 2D $hole$ system (2DHS) but there they flanked the $\nu=1/3$ FQHS \cite{Santos.PRL.1992, Santos.PRB.1992}. Another system, where insulating phases next to the $\nu=1/3$ FQHS have been reported, is the ZnO 2DES \cite{Maryenko.NatComm.2018}. These reports provided experimental evidence for the importance of LL mixing (LLM) in favoring the WS states over the FQH liquid states, and moving the stability of the WS to fillings as large as $\simeq 1/3$. Note that the 2D holes in GaAs and 2D electrons in ZnO have an effective mass ($m^*$) which is larger than $m^*$ of GaAs 2D electrons by a factor of $\sim6$ \cite{Tsukazaki.PRB.2008}, leading to a larger LLM parameter $\kappa$, defined as the ratio of the Coulomb energy and the LL separation: $\kappa=(e^2/4 \pi \epsilon_0 \epsilon l_B)/(\hbar eB/m^*)$, where $l_B=\sqrt{\hbar /eB}$ is the magnetic length, and $\epsilon$ is the dielectric constant. The importance of LLM in the competition between the liquid and solid states at low $\nu$ has indeed been highlighted by a number of theoretical \cite{Yoshioka.JPSP.1984,Yoshioka.JPSP.1986,Zhu.PRL.1993,Price.PRL.1993,Platzman.PRL.1993,Ortiz.PRL.1993,Zhao.PRL.2018,Zuo.PRB.2020} and additional experimental studies \cite{Li.PRL.1997,Li.PRB.2000,Csathy.PRL.2004,Csathy.PRL.2005,Pan.PRB.2005,Knighton.PRB.2018,Ma.PRL.2020,Bayot.EuroPhys.1994}.

\begin{figure*}[t!]
  \centering
    \psfig{file=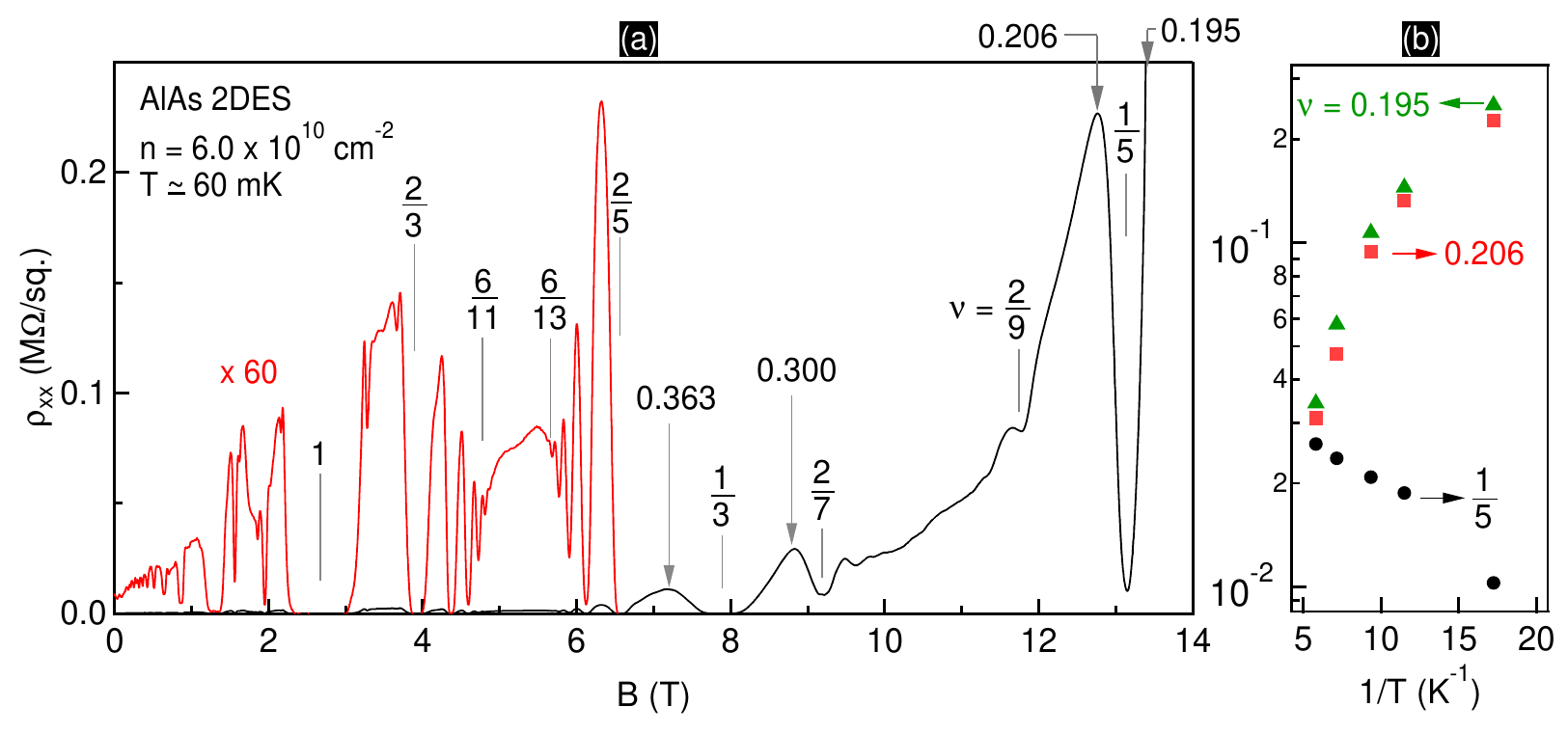, width=1\textwidth}
  \centering
  \caption{\label{transport}
   (a) Longitudinal resistivity $\rho_{xx}$ vs. perpendicular magnetic field $B$ for an AlAs 2DES with $n = 6.0 \times10^{10}$ cm$^{-2}$ at \textit{T} $\simeq60$ mK. The red trace is expanded by a factor of 60. Numerous FQHSs, marked by vertical lines, attest to the extremely high quality of the sample. (b) Arrhenius plots of $\rho_{xx}$ at $\nu = 0.206$, $1/5$, and $0.195$. The insulating phases on the flanks of $\nu=1/5$ signal a disorder-pinned Wigner solid.
  }
  \label{fig:transport}
\end{figure*}

Here we report the observation of FQH and WS phases in 2DESs confined to AlAs quantum wells (QWs). These samples, which have exceptionally high quality, provide an ideal platform to explore the liquid-solid transitions for a large range of $\kappa$ and $\nu$. Electrons in AlAs QWs have larger $m^{*}$ ($\simeq0.46$ in units of free electron mass) and smaller $\epsilon$ ($\simeq10$) compared to GaAs 2DESs ($m^{*}\simeq0.067$ and $\epsilon\simeq13$). As a result, the $\kappa$ parameter for AlAs 2DESs, for a given density and filling factor, is $\simeq9$ times larger than in GaAs 2DESs. Furthermore, the LLs are simple and their energies vary linearly with $B$, making the interpretation of $\kappa$ straightforward. This is in contrast to the much more complex LLs in the 2DHS which are non-linear and have crossings as a function of $B$ \cite{Ma.PRL.2020,Winkler.SpinOrbit.2003}. Our data provide a $\kappa$ vs. $\nu$ phase diagram for the liquid-solid states in a large range of fillings ($1/5 \lesssim \nu \lesssim 1/3$) and LLM ($4 \lesssim \kappa \lesssim 12$); note that the typical value for $\kappa$ in GaAs 2DESs is $\simeq1$. We compare this diagram quantitatively to the theoretically available diagrams near $\nu=1/3$ and $1/5$ \cite{Zhao.PRL.2018}. The overall agreement is quite good and demonstrates the role of LLM, \textit{as well as} finite-layer thickness; however, some subtle discrepancies remain. We also report a thermal melting phase diagram, near $\nu=1/3$, for the pinned WS.

We studied 2DESs confined to modulation-doped AlAs QWs grown on GaAs (001) substrates. In the absence of symmetry-breaking, in-plane strain, our samples, which have QW widths (\textit{w}) ranging from $20$ to $50$ nm, host electrons occupying two anisotropic in-plane valleys. The valleys have longitudinal and transverse effective masses $m_{l}=1.1$ and $m_{t}=0.20$, leading to an effective cyclotron mass $m^{*}= (m_{l}\times m_{t})^{1/2}=0.46$ \cite{Shayegan.review.2006}. In our experiments, thanks to either intentionally applied uniaxial strain, or unintentional residual strain during the cooldown, the electrons occupy only one valley. The samples have 2DES densities ($n$) ranging from $3.3$ to $33$, in units of $10^{10}$ cm$^{-2}$, which we will use throughout the manuscript. They have a van der Pauw geometry, with alloyed InSn contacts at the corners and edge midpoints. We made measurements primarily in a dilution refrigerator, and used low-frequency ($\leq 7$ Hz) lock-in techniques: constant-current mode for magnetoresistance and constant-voltage mode for differential resistance. We provide further details of sample parameters and device fabrication in the Supplemental Materials (SM) \cite{noauthor_supplementary_nodate}.

Figure 1(a) shows the longitudinal resistance ($\rho_{xx}$) vs. $B$ at \textit{T} $\simeq60$ mK for a sample with \textit{n} $= 6.0$. The expanded red trace shows several FQHSs, manifested by deep minima in the $\rho_{xx}$ trace. At high magnetic fields, the $\nu = 1/5$ FQHS exhibits a strong minimum. The numerous FQHSs and especially the $\nu=1/5$ state, which has only been observed in the highest quality GaAs 2DESs \cite{Jiang.PRL.1990,Goldman.PRL.1990,Shayegan.Review.Das.Sarma&Pinzuk.Book.1996,Pan&Tsui.PRL.2002} and more recently in graphene samples \cite{Zeng&Dean.PRL.2019,Zhou.NatPhys.2020}, attest to the exceptionally high quality of the sample. On the flanks of $\nu=1/5$ (e.g., at $\nu=0.206$ and $0.195$), $\rho_{xx}$ has very high values. Figure 1(b) shows Arrhenius plots of $\rho_{xx}$ at $\nu=1/5$ and its flanks. At $\nu=1/5$, $\rho_{xx}$ decreases as we lower the temperature, consistent with the formation of a FQHS. In contrast, on its flanks, $\rho_{xx}$ increases by about one order of magnitude when we lower the temperature from $200$ to $60$ mK.  Such insulating phases are generally believed to signal a disorder-pinned WS state \cite{Jiang.PRL.1990, Goldman.PRL.1990, Jiang.PRB.1991, Santos.PRB.1992, Santos.PRL.1992, Williams.PRL.1991, Sajoto.Thesis.1993, Li.PRL.1997, Paalanen.PRB.1992,Maryenko.NatComm.2018}. Our data and analysis, as we describe below, are consistent with this interpretation. Moreover, they allow us to determine a quantitative quantum melting phase diagram for the magnetic-field-induced WS at low fillings, and compare this diagram to a recently calculated diagram, as reproduced in Fig. 2.

Figure 2 summarizes our experimental results (symbols) in $\kappa$ vs. $\nu$ phase diagrams near $\nu=1/3$ and $1/5$. The grey lines and yellow/white regions are from theoretical calculations \cite{Zhao.PRL.2018}. The experimental data are shown by symbols. The open, closed, and semi-open symbols represent a liquid, a disorder-pinned WS, and a close competition between liquid-solid, respectively. Data for \textit{n} $=6.0$ are shown in Fig. 2 by red symbols. Around $\nu=1/5$, in Fig. 2(b), there is overall agreement with the theoretical calculations except for $\nu=2/9$. At $\nu=2/9$, as the temperature decreases, $\rho_{xx}$ increases, yet it remains a local minimum (see Fig. 1(a)). This behavior suggests a competition between the FQH liquid and a disorder-pinned WS. It is likely that disorder is partly responsible for the competition. It is worth remembering that early GaAs 2DESs, which had lower quality, showed a competition between a FQHS and an insulating behavior at $\nu=1/5$ \cite{Willett.PRB.1988}. When samples of much better quality became available, $\rho_{xx}$ at $\nu=1/5$ displayed a clear FQHS with a vanishing resistance at the lowest temperatures \cite{Jiang.PRL.1990}.
Also, previous experimental studies in samples with a controlled (minute) amount of alloy disorder report an insulating behavior at and around $\nu=1/5$ \cite{Li.PRL.2010,Moon.PRB.2014}. Local compressibility measurements in GaAs 2DES showed that disorder creates puddles of slightly different densities, and as a result of $\Delta\nu$ \cite{Ilani.Nature.2004,Venkatachalam.Nature.2011}. In the theoretical phase diagram reproduced in Fig. 2, at large $\kappa$, variations in $\nu$ would make the WS (yellow region) overcome the $2/9$ FQH liquid (narrow white region), preventing its percolation. For a detailed discussion of the role of disorder in the competition between FQHSs and solids, see Ref. \cite{Zuo.PRB.2020}.

\begin{figure}[t!]
  \centering
    \psfig{file=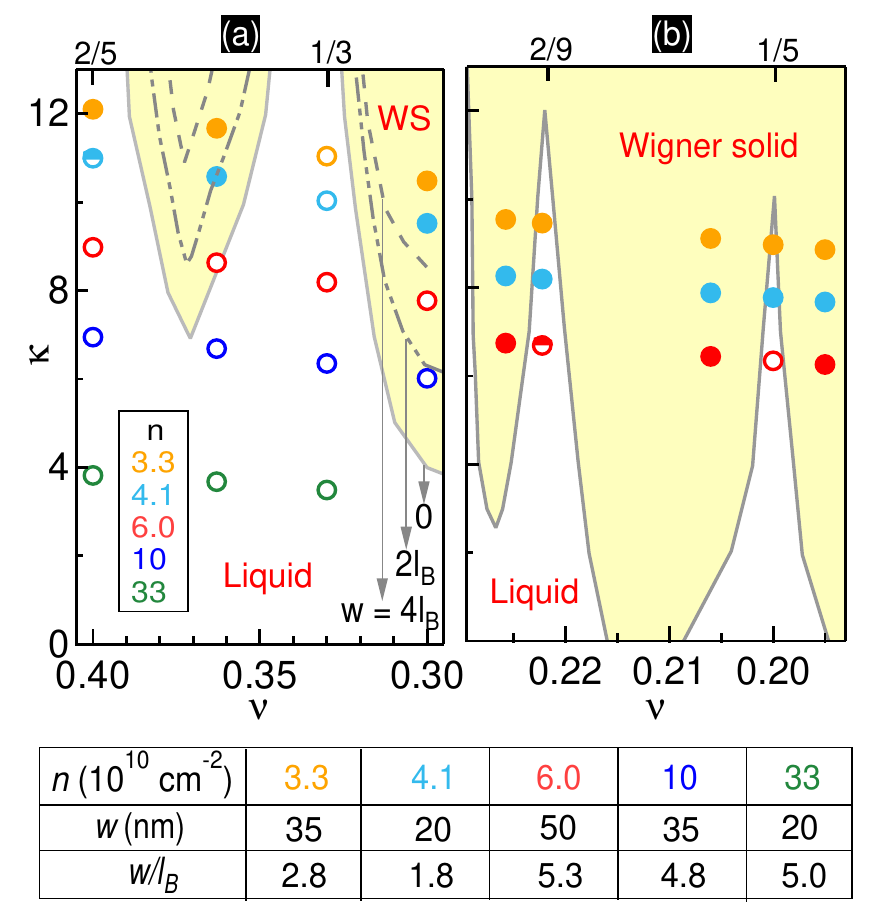, width=0.48\textwidth}
  \centering
  \caption{\label{Ip}
   Liquid-Wigner solid \textit{quantum melting} phase diagrams around (a) $\nu=1/3$, and (b) $\nu=1/5$; $\kappa$ is the Landau level mixing parameter. The grey solid lines represent zero-layer-thickness calculations by Zhao \textit{et al.} \cite{Zhao.PRL.2018} for the boundary between the WS and the liquid phases. The closed and open circles indicate the experimentally-deduced WS and liquid phases, respectively. The half-filled circles are used to imply a close competition between WS and liquid phases. The filling factors at which the experimental data are presented are $2/5$, $0.363$, $1/3$, $0.300$, $2/9$, $0.206$, $1/5$, and $0.196$. In (a), the dotted-dashed and dashed lines represent the theoretical boundaries between the liquid and WS states for \textit{w} $=2l_{B}$ and \textit{w} $=4l_{B}$, respectively \cite{Zhao.PRL.2018}. The table shows the density, QW width, and \textit{w}$/l_{B}$ (at $\nu=1/3$) for the different samples that we studied. 
  }
  \label{fig:Ip}
\end{figure}

The red symbols in Fig. 2(a) show our experimental results for \textit{n} $=6.0$ near $\nu\simeq1/3$. The $\rho_{xx}$ trace (see Fig. 1(a)) exhibits fully-developed $\nu=1/3$ and $2/5$ FQHSs, which agree with the prediction of liquid phases (open red symbols in Fig. 2(a)). As seen in Fig. 1(a), $\rho_{xx}$ maxima on the flanks of $\nu=1/3$ (at $\nu=0.363$ and $0.300$) are several times larger than the low-field resistance, but do not show strong insulating behavior. This behavior appears to be at odds with the calculations. However, the calculations that delimit the yellow/white regions in Fig. 2 are for a 2DES with zero thickness, but our 2DES is confined to a finite-width QW. Zhao \textit{et al.} \cite{Zhao.PRL.2018} reported that for a 2DES with finite layer thickness, the WS-liquid boundary moves to larger values of $\kappa$ (see SM of Ref. \cite{Zhao.PRL.2018}). In Fig. 2(a), we also include the results of finite-thickness calculations from Ref. \cite{Zhao.PRL.2018}, for \textit{w} $=2l_{B}$ and $4l_{B}$. As seen in the table shown in the lower part of Fig. 2, our AlAs 2DES with \textit{n} $=6.0$ has \textit{w}$/l_{B}\simeq5.3$ at $\nu=1/3$. Comparing the experimental data with the theoretical \textit{w} $=4l_{B}$ liquid-solid boundary, we find agreement with the liquid states predicted in the calculations.

\begin{figure}[b!]
  \centering
    \psfig{file=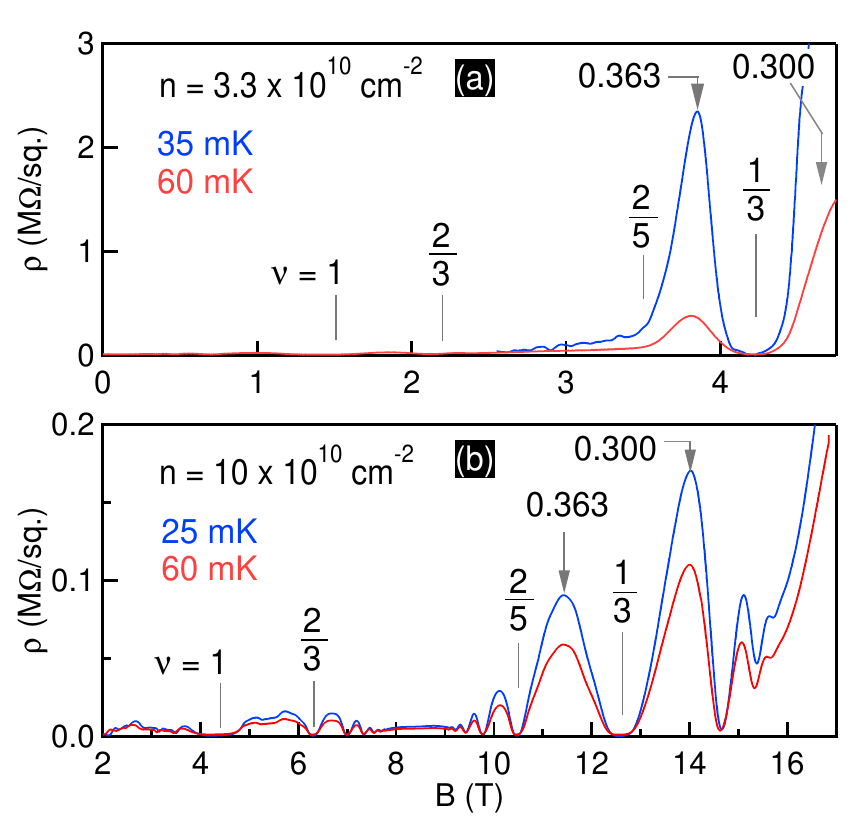, width=0.48\textwidth}
  \centering
  \caption{\label{Ip}
  $\rho_{xx}$ vs. $B$ for AlAs 2DESs with (a) $n = 3.3 \times10^{10}$ cm$^{-2}$, and (b) $n = 10 \times10^{10}$ cm$^{-2}$.
  }
  \label{fig:Ip}
\end{figure}

In order to explore larger values of $\kappa$, we studied samples with densities \textit{n} = $4.1$ and $3.3$ (light-blue and yellow symbols in Fig. 2, respectively). Figure 3(a) shows $\rho_{xx}$ vs. $B$ at \textit{T} $\simeq35$ and $60$ mK for \textit{n} $=3.3$ ($\rho_{xx}$ for \textit{n} $=4.1$ has similar characteristics, see SM). On the flanks of the $\nu=1/3$ FQHS, $\rho_{xx}$ has very large values and exhibits a strong insulating behavior, which we interpret as a signature of a disorder-pinned WS. The samples with \textit{n} $=3.3$ and $4.1$ have \textit{w}$/l_{B}\simeq2.8$ and $1.8$, respectively. When compared to the solid-liquid calculated boundary with \textit{w} $=2l_{B}$, we find agreement with the predicted WS on the flanks of $1/3$. The behavior at  $\nu=2/5$ is an exception to the agreement: in the experiments we find a WS for $n=3.3$, and a competition between a FQHS and a solid for \textit{n} $=4.1$.  For smaller $\nu$, around $1/5$ in Fig. 2(b), there is a strong insulating phase without any hint of FQHSs. As discussed in the previous paragraph, we invoke the role of disorder to explain the above discrepancies with calculations.

We also studied samples with densities \textit{n} = $10$ and $33$, to explore smaller values of $\kappa$. The data are shown in Fig. 2 by dark-blue and green symbols, respectively. Figure 3(b) shows $\rho_{xx}$ vs. $B$ at \textit{T} $\simeq25$ and $60$ mK for the sample with \textit{n} $=10$. We find fully-developed FQHSs at $\nu=1/3$ and $2/5$. At $\nu=0.363$ and $0.300$, $\rho_{xx}$ does not show a strong insulating behavior, suggesting that there is no WS on the flanks of $\nu=1/3$. Considering that the sample with \textit{n} $=10$ has \textit{w}$/l_{B}\simeq4.8$ at $\nu=1/3$, the experimental data points agree with the theoretical boundary, once the finite-thickness of the 2DES is taken into account. Note that the filling range near $\nu=1/5$ was experimentally out of our reach for \textit{n} $=10$. At \textit{n} $=33$, the experimental data near $\nu=1/3$ are in good overall agreement with calculations (the $\rho_{xx}$ trace is shown in the SM); there is no sign of an insulating phase down to $\nu\simeq0.32$, the smallest $\nu$ we could reach experimentally for this sample in a hybrid magnet.

Two remarks are in order here. First, the WS and FQH states near $\nu=1/5$ are very close in energy, and the role of finite-layer-thickness could not be assessed reliably in calculations \cite{Zhao.PRL.2018}. Second, as seen in Fig. 2(b), the range of $\nu$ near $\nu=1/5$ (as well as near $\nu=2/9$) where the FQHS is stable, at large $\kappa$, is extremely narrow according to the calculations. This may explain why FQHSs are not experimentally seen for the lower density samples: even a very small level of disorder can cause density inhomogeneity (of the order of 1$\%$), which can lead to the disappearance of the $\nu=1/5$ FQHS. In this context, our observation of a well-developed $\nu=1/5$ FQHS, with a vanishing $\rho_{xx}$ as the temperature approaches zero, in the $n=6.0$ sample (Fig. 1) at $\kappa$ exceeding 6 (Fig. 2(b)) attests to the exceptionally high quality and homogeneity of the sample.



\begin{figure}[t!]
  \centering
    \psfig{file=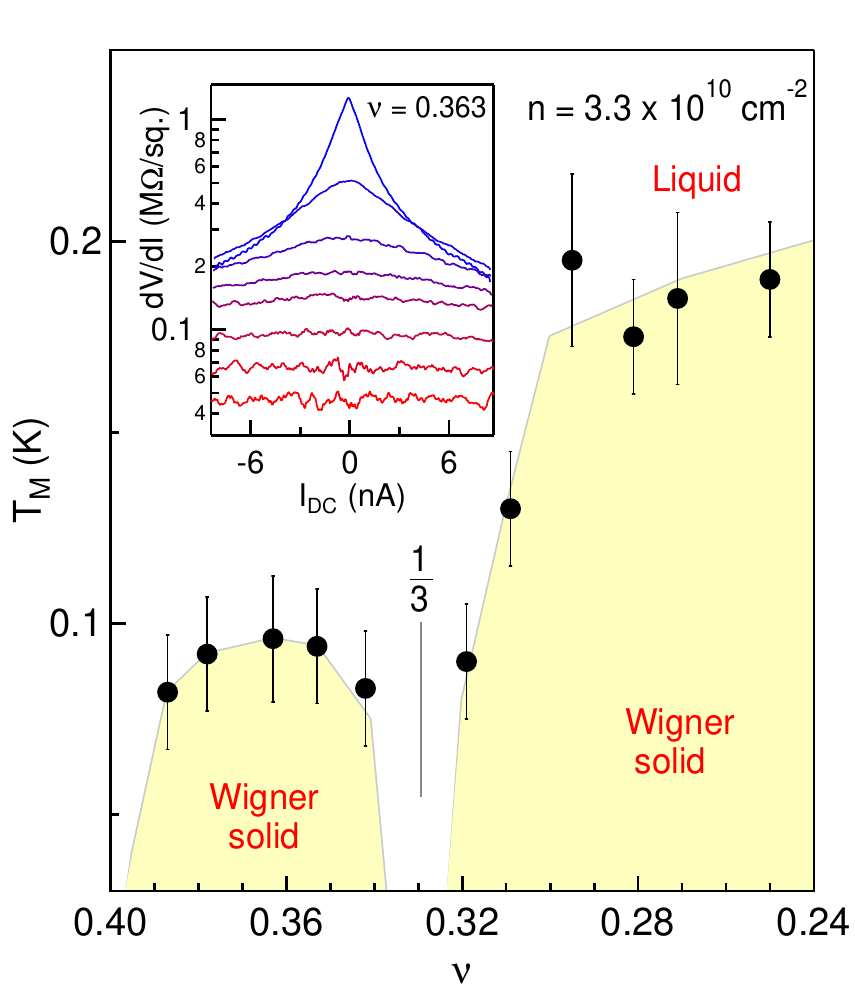, width=0.45\textwidth}
  \centering
  \caption{\label{Ip}
Wigner solid \textit{thermal melting} phase diagram. The light yellow and white regions indicate the solid and liquid phases, respectively. The grey line is a guide to the eye. The inset shows the temperature dependence of the differential resistance $dV/dI$ vs. $I_{DC}$ traces at $\nu = 0.363$. The temperatures are (from top to bottom): $40$, $53$, $67$, $80$, $90$, $113$, $145$, and $206$ mK.
  }
  \label{fig:phase2}
\end{figure}

Another property of interest for the WS is its \textit{thermal melting} phase diagram \cite{Ma.PRL.2020,Deng&Shayegan.PRL.2019,Goldman.PRL.1990,Jiang.PRB.1991,Chen&Engel.NatPhys.2006,Williams.PRL.1991,Sajoto.Thesis.1993,Shayegan.Review.Das.Sarma&Pinzuk.Book.1996,Bayot.EuroPhys.1994}. To probe this thermal melting, we monitor the non-linear behavior of the differential resistance ($dV/dI$) as a function of a DC current ($I_{DC}$) flowing through the 2DES with \textit{n} = $3.3$. This technique was used in the past to study the magnetic-field-induced WS in other 2D carrier systems \cite{Goldman.PRL.1990,Williams.PRL.1991,Santos.PRL.1992,Sajoto.Thesis.1993,Shayegan.Review.Das.Sarma&Pinzuk.Book.1996}. Measurement details are outlined in the SM \cite{noauthor_supplementary_nodate}.

Figure 4 inset shows the temperature dependence of $dV/dI$ vs. $I_{DC}$ at $\nu = 0.363$. At the lowest temperatures, the 2DES shows increased conduction for $I_{DC}\neq0$, signaled by a decrease in $dV/dI$, and consistent with a disorder-pinned WS \cite{Williams.PRL.1991,Jiang.PRB.1991,Sajoto.Thesis.1993,Goldman.PRL.1990,Santos.PRL.1992,Shayegan.Review.Das.Sarma&Pinzuk.Book.1996}. At the highest temperatures, where we expect the WS to have melted, $dV/dI$ is independent of $I_{DC}$, consistent with ohmic conduction observed in liquid phases \cite{Williams.PRL.1991,Jiang.PRB.1991,Sajoto.Thesis.1993,Goldman.PRL.1990,Santos.PRL.1992,Santos.PRL.1992,Knighton.PRB.2018,Maryenko.NatComm.2018,Shayegan.Review.Das.Sarma&Pinzuk.Book.1996}, and in our sample very near the $\nu=1/3$ FQHS (data not shown). As we raise the temperature, there is a gradual change from the low-temperature, non-linear behavior to the high-temperature, linear limit. This temperature dependence is qualitatively generic for all the filling factors where we observe a strong insulating behavior in this sample, namely near $\nu\simeq0.36$ and for $\nu \lesssim 0.32$. We deduce the melting temperature ($T_{M}$) of the WS by estimating the temperature at which $dV/dI$ changes from non-linear to ohmic; see SM for details.  

Figure 4, which contains $T_{M}$ vs. $\nu$, displays our deduced WS thermal melting phase diagram of the AlAs 2DES with $n = 3.3$. As $\nu$ approaches $0.33$, $T_{M}$ decreases sharply and vanishes close to the $\nu=1/3$ FQHS. When $\nu$ is larger than $1/3$, we observe a reentrant WS. The values of $T_{M}$ in our thermal phase diagram are close to those recently reported for the melting temperatures of a WS reentrant around $\nu=1/3$ in a GaAs 2D hole system of density $3.8$ \cite{Ma.PRL.2020}.

Our experimental data and their quantitative comparison with state-of-the-art calculations provide strong evidence for the validity of associating the insulating phases, observed in ultra-high-quality 2DESs, at low fillings, on the flanks of FQHSs with disorder-pinned WS states. The data also highlight the importance of LLM, as well as the finite-layer-thickness, in the competition between the correlated solid-liquid phases. The role of finite-layer-thickness was not assessed experimentally prior to our study. It is also important to note that the calculations \cite{Zhao.PRL.2018} were done for isotropic 2DESs while our data are taken on samples which have an effective mass anisotropy of $\simeq 5$ \cite{Shayegan.review.2006}. The very good agreement between the calculations and the experimental data might thus sound surprising. However, as reported experimentally \cite{Jo&Shayegan.PRL.2017} and theoretically \cite{Wang&Zhang.PRB.2012,Qiu&Haldane.PRB.2012,Yang&Haldane.PRB.2013,Yang.PRB.2013,Johri&Haldane.NJP.2016,Balram&Jain.PRB.2016}, it turns that FQHSs are quite robust against anisotropy, and that (anisotropic) WS states only become prevalent in systems with much larger anisotropies than in our samples. Finally, we emphasize that our data described here were taken in AlAs 2DESs with only one conduction-band valley occupied. An exciting future study would be to experimentally and/or theoretically assess the stability of liquid and solid states in AlAs 2DESs where the electrons occupy two degenerate valleys.

\begin{acknowledgments}

We acknowledge support by the DOE BES (Grant No. DE-FG02-00-ER45841) for measurements, and the NSF (Grants No. DMR 1709076, No. ECCS 1906253, and No. MRSEC DMR 1420541), and the Gordon and Betty Moore Foundation's EPiQS Initiative (Grant No. GBMF9615 to L.N.P.) for sample fabrication and characterization. This research is funded in part by QuantEmX grants from Institute for Complex Adaptative Matter. A portion of this work was performed at the National High Magnetic Field Laboratory, which is supported by National Science Foundation Cooperative Agreement No. DMR-1644779 and the state of Florida. We thank S. Hannahs, T. Murphy, A. Bangura, G. Jones, and E. Green at NHMFL for technical support. We also thank J. K. Jain for illuminating discussions.

\end{acknowledgments}


\end{document}


\setcounter{page}{1}

\title[]{Supplementary Material for "Competition between fractional quantum Hall liquid and Wigner solid at small fillings: Role of layer thickness and Landau level mixing"}
\author{K. A. \surname{Villegas Rosales}}
\author{S. K. \surname{Singh}}
\author{Meng K. \surname{Ma}}
\author{Shafayat M. \surname{Hossain}}
\author{Y. J. \surname{Chung}}
\author{L. N. \surname{Pfeiffer}}
\author{K. W. \surname{West}}
\author{K. W. \surname{Baldwin}}
\author{M. \surname{Shayegan}}
\affiliation{Department of Electrical Engineering, Princeton University, Princeton, New Jersey 08544, USA}
\date{\today}

\maketitle

\section{Magnetoresistance data at additional densities}

\begin{figure*}[h!]
  \centering
    \psfig{file=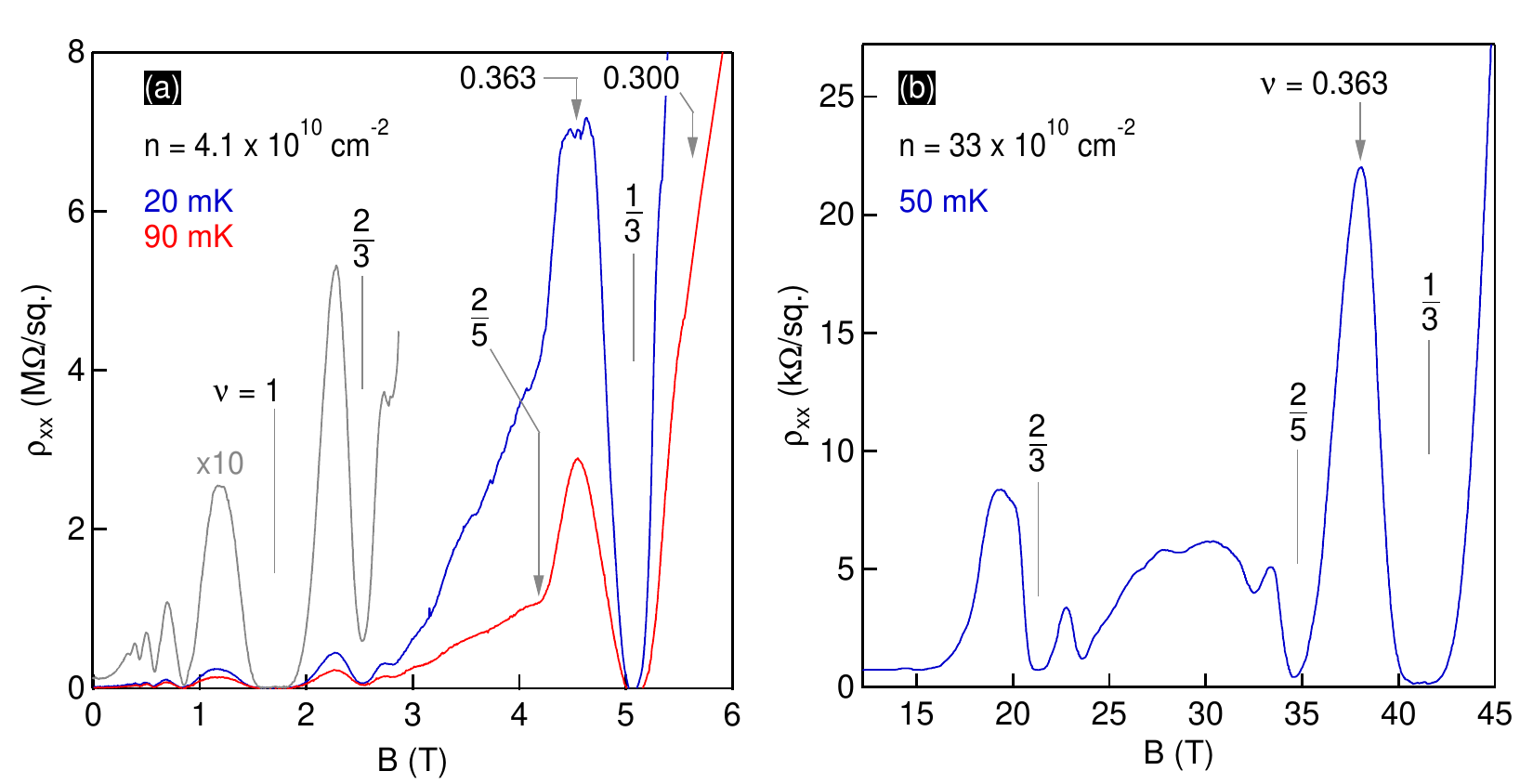, width=1\textwidth}
  \centering
  \caption{\label{transport}
   Longitudinal resistivity $\rho_{xx}$ vs. perpendicular magnetic field $B$ for an AlAs 2DES with (a)  $n = 4.1 \times10^{10}$ cm$^{-2}$, and (b) $n = 33 \times10^{10}$ cm$^{-2}$. The grey trace is expanded by a factor of 10 in Fig. S1(a).
  }
  \label{fig:transport}
\end{figure*}

Figure S1 shows the longitudinal resistivity ($\rho_{xx}$) vs. perpendicular magnetic field ($B$) for densities (a) \textit{n} $=4.1\times10^{10}$ cm$^{-2}$, and (b) \textit{n} $=33\times10^{10}$ cm$^{-2}$. The $\rho_{xx}$ traces correspond to the light-blue and green symbols in Fig. 2 (see main text), respectively. In Fig. S1(a), the $\nu=1/3$ fractional quantum Hall state (FQHS) is flanked by highly-resistive regions, on the order of $\simeq7$ M$\Omega$. As we raise the temperature \textit{T} from $20$ to $90$ mK, $\rho_{xx}$ at $\nu=0.363$ decreases $\simeq50\%$. At $\nu=2/5$, at \textit{T} $\simeq90$ mK, the $\rho_{xx}$ trace shows a weak minimum. We infer from the experimental data that the ground state at $\nu=1/3$ is a FQH liquid, while the insulating behavior at $\nu=0.363$ and $0.300$ suggest pinned Wigner solid states. The trace in Fig. S1(b) was measured in a hybrid magnet at the Naiontal High Magnetic Field Laboratory in Tallahassee, Fl. The overall value of $\rho_{xx}$ is on the order of $10$ k$\Omega$, and the 2DES at $\nu=2/5$, $0.363$, and $1/3$ behaves as a liquid.

\newpage

\section{Differential resistance data of an AlAs 2DES in the disorder-pinned Wigner solid region}

\begin{figure*}[h!]
  \centering
    \psfig{file=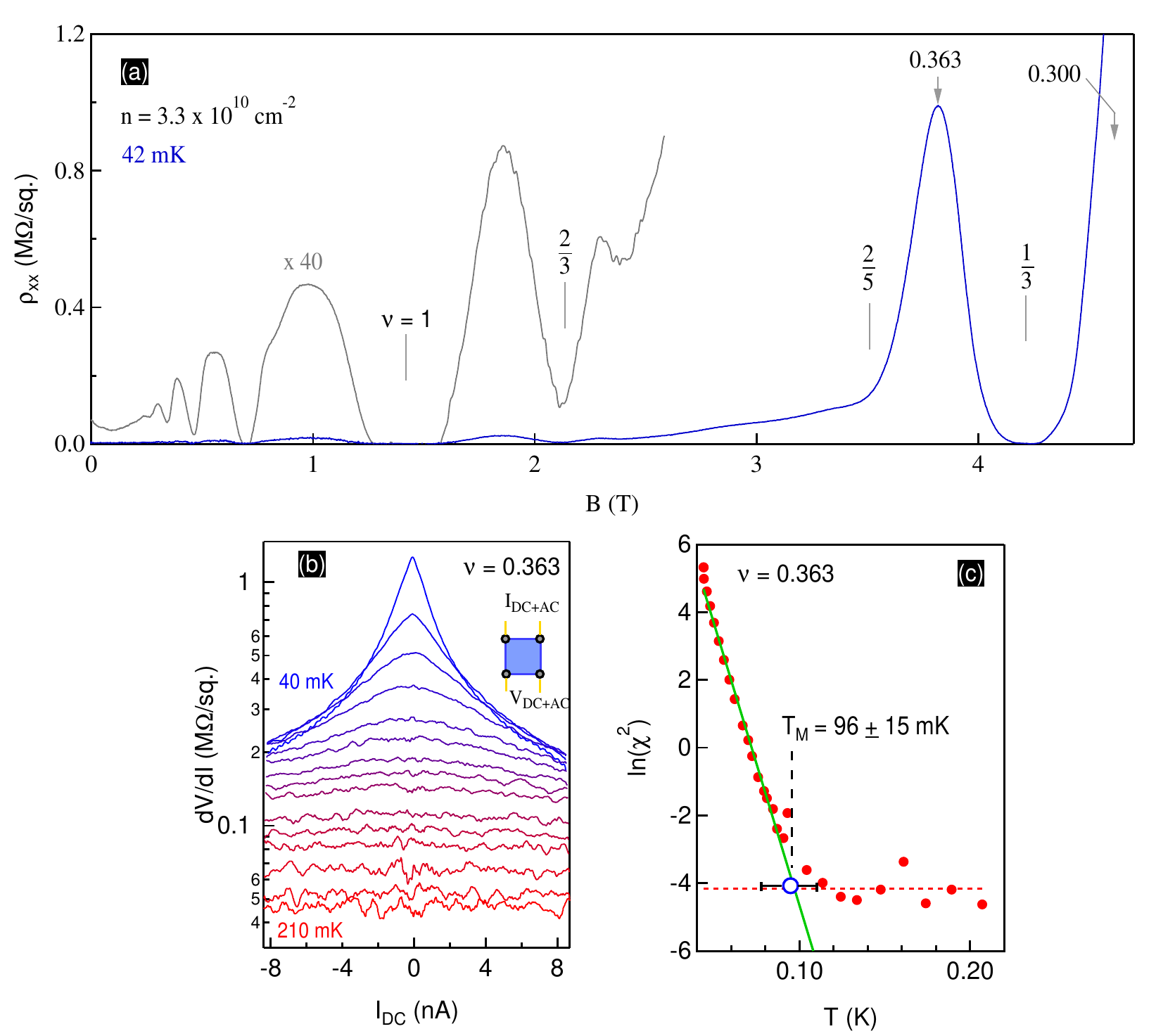, width=0.9\textwidth}
  \centering
  \caption{\label{transport}
   (a) $\rho_{xx}$ vs. $B$ for an AlAs 2DES with \textit{n} $=3.3\times10^{10}$ cm$^{-2}$ at \textit{T} $\simeq42$ mK. (b) Temperature dependence of the differential resistance ($dV/dI$) vs. DC current ($I_{DC}$) at $\nu=0.363$. The inset shows a schematic of the device. We performed AC and DC measurements in a 4-terminal configuration. (c) Natural logarithm of $\chi^{2}$ vs. \textit{T} for $\nu=0.363$. The melting temperature ($T_{M}$) is deduced from the intersection of the green and red lines.
  }
  \label{fig:transport}
\end{figure*}

Figure S2 shows details of the differential resistance ($dV/dI$) vs. DC current ($I_{DC}$) measurements in the disorder-pinned Wigner solid. In Fig. S2(a), the $\rho_{xx}$ values flanking the $\nu=1/3$ FQHS are very large in comparison to the low-field values. For a comparison we expanded the low-field $\rho_{xx}$ by 40 times (grey trace in Fig. S2(a)). The highly-resistive regions at $\nu=0.363$ and $0.300$ indicate a disorder-pinned Wigner solid.

The Wigner solid, rendered immobile by the residual impurities, can be made to conduct electricity by the application of an external $I_{DC}$. More specifically, we apply AC ($0.21$ Hz and $2$ mV) and DC (up to $\pm 100$ mV) excitation voltages through a $10$ M$\Omega$ resistor to a pair of contacts (see inset in Fig. S2(b)). The AC voltage ($V_{AC}$) allows us to pick up $dV/dI$ in another pair of contacts (four-terminal measurement). The DC voltage ($V_{DC}$) generates a DC current ($I_
{DC}$) flowing across the device. The magnitude of $dV/dI$ is then probed in response to $I_{DC}$.

In Fig. S2(c) we illustrate our procedure for a quantitative determination of the Wigner solid melting temperature ($T_{M}$). First, we do linear fits to the $dV/dI$ traces, at different $T$, and deduce the $\chi^{2}$ values. At the lowest temperature the $dV/dI$ trace is very non-linear and $\chi^{2}$ is very large. As we raise the temperature $\chi^{2}$ decreases. In the high-temperature limit, $\chi^{2}$ oscillates around a very small constant value, which shows that $dV/dI$ is linear, i.e., that transport is Ohmic. The very small value of $\chi^{2}$ at high temperatures results from the small, but finite, noise in our measurements; see the high-\textit{T} traces in Fig. S2(b). We take the natural logarithm of $\chi^{2}$ (ln($\chi^{2}$)) and plot it against \textit{T}, as shown in Fig. S2(c). The green line is a linear fit to the ln($\chi^{2}$) in the low-temperature regime, and the red-dashed line is a fit to the oscillating values in high-temperature limit. We deduce $T_{M}$ from the intersection of these two lines.